\documentclass[12pt]{article}

\usepackage{amsmath,amssymb}
\usepackage{natbib}
\usepackage{enumerate}
\usepackage{graphicx}
\usepackage{longtable}
\usepackage{subcaption}
\usepackage{amsthm}
\usepackage{array}
\usepackage{float}
\usepackage[utf8]{inputenc}
\usepackage[inline]{trackchanges} 
\usepackage[affil-it]{authblk}
\usepackage[margin=.7in]{geometry}

\graphicspath{ {figs/} }
\addeditor{Alan}
\addeditor{Wilson}
\addeditor{Nima}

\theoremstyle{remark}

\begin{document}

\title{Data-adaptive statistics for multiple hypothesis testing in
high-dimensional settings}

\author{Weixin Cai
}
\author{Nima S. Hejazi
}
\author{Alan E. Hubbard
}
\affil{Division of Biostatistics, University of California, Berkeley}

\date{\today}
\maketitle

\begin{abstract}

Current statistical inference problems in areas like astronomy, genomics, and
marketing routinely involve the simultaneous testing of thousands -- even
millions -- of null hypotheses. For high-dimensional multivariate
distributions, these hypotheses may concern a wide range of parameters, with
complex and unknown dependence structures among variables. In analyzing such
hypothesis testing procedures, gains in efficiency and power can be achieved by
performing variable reduction on the set of hypotheses prior to testing. We
present in this paper an approach using data-adaptive multiple testing that
serves exactly this purpose. This approach applies data mining techniques to
screen the full set of covariates on equally sized partitions of the whole
sample via cross-validation. This generalized screening procedure is used to
create average ranks for covariates, which are then used to generate a reduced
(sub)set of hypotheses, from which we compute test statistics that are
subsequently subjected to standard multiple testing corrections. The principal
advantage of this methodology lies in its providing valid statistical inference
without the \textit{a priori} specifying which hypotheses will be tested.
Here, we present the theoretical details of this approach, confirm its validity
via simulation study, and exemplify its use by applying it to the analysis of
data on microRNA differential expression.

\vspace{9pt}
\par
\end{abstract}

\newpage

\section{Introduction}\label{intro}

Recently developed technologies enable high throughput screen of thousands (millions) of biological molecules, which has resulted in filters that use multiple testing to highlight potentially informative biomakers.  For instance, consider microRNA, a new class of small, non-coding RNAs have been the subject
of intense study due to their central role in gene regulation
\citep{wienholds2005microrna}.  In a process involving binding to messenger RNA
(mRNA), miRNAs regulate gene expression at the post-transcriptional level,
thereby affecting the abundance of a wide range of proteins in diverse
biological processes.   The resulting data consists of a large vector of microRNA expressions as well as other characteristics and experimental conditions relevant to a particular biological sample. Before looking at the complex relationship of these molecules, a first step is often examining the association of microRNA expression with some other phenotypic variable(s).  In our case, we consider a study examining the relationship of occupational exposure benzene (a known carcinogen) to  microRNA expression. This study, like many of its kind, has a relative small sample size, but large ambitions regarding the teasing out of associations of many thousands of potential microRNA's.  In this paper, we propose a method that adaptively reduces the number of tests so that a study can preserve reasonable power even when the number of potential tests is huge.

Speaking generally, current problems of statistical inference involving multiple
hypothesis testing share the following characteristics: inference for high-
dimensional multivariate distributions, with complex and unknown dependence
structures among variables; a variety of parameters of interest, such as
coefficients in general regression models relating possibly censored biological
and clinical covariates and outcomes to genome-wide expression measures and
genotypes; many null hypotheses, in the thousands or even millions; complex and
unknown dependence structures among test statistics.

An often-ignored yet insidious issue with small-sample inference and large
numbers of comparisons is the enormous sample sizes required for joint
convergence, by the central limit theorem, of the joint sampling distribution
to a multivariate normal
distribution. Unfortunately, others have shown that for convergence to multivariate normal distribution
in the far tails (to derive accurate adjustment for multiple testing) requires astronomical
sample sizes  (\textit{e.g.}, from high-throughput sequencing technologies).
\cite{gerlovina2016big} employed Edgeworth expansions to rigorously study the
manner in which the number of tests performed affects the accuracy of the
resultant analyses, with particular attention paid to situations where the
number of tests vastly outnumbers the sample size. Here, the utility of
Edgeworth series lies in their providing, by way of higher-order
approximations, estimates of critical values that would otherwise be computed
exactly were the true distribution known.  In any case, these expansions and associated simulations
of multiple testing experiments showed that error control can be wildly anti-conservative if sample size is
inadequate.   Thus, using
standard and commonly used multiple testing techniques can make it practically
impossible to obtain honest statistical inference when conducting very large
numbers of tests.

Motivated by the broad use of multiple testing procedures for very large
numbers of tests, and the limitations of existing multiple testing procedures,
we present in this paper a technique for data-adaptive multiple testing in
high-dimensional problems, one that harnesses data mining procedures to perform
variable reduction, whilst preserving accurate and honest statistical
inference. This new method is a natural extension of the data-adaptive
statistical target parameter framework introduced by \cite{hubbard2016mining},
who show that this new class of inference procedures provides an impetus for
using methods providing rigorous statistical inference for data-mining procedures.

The proposed approach for multiple testing in high-dimensional settings uses
data-adaptive test statistics, which rely on cross-validation, to perform
variable reduction by screening algorithms.  Typically, the method uncovers
associations that represent signals that are stable across the full sample,
while allowing for multiple testing efforts to be restricted to a much smaller
subset of biomarkers (predictors). It is expected that this new class of test statistics
outperforms, in terms of improvements in both power and control of type-I error
rate, standard test statistics generated by classical approaches to multiple
hypothesis testing, in which variable set is  pre-specified.

In  section \ref{methods}, we proceed to define this methodology in generality
with the aforementioned estimation strategies, present theorems establishing its
asymptotic statistical performance, present influence function and
bootstrap-based inference, and discuss the implications of these theoretical
results. In section \ref{sim}, we demonstrate the relative performance by way of simulation studies;
in section \ref{mirna}, we apply the general approach for high-dimensional multiple testing using
data-adaptive test statistics to the analysis of data from miRNA assays; and
then conclude (in section \ref{discussion}) with remarks on the implications of this new methodology to
future work in analyzing data from biomedical investigations.

\newpage

\section{Methodology}\label{methods}

\subsection{Notation and setup}\label{notation}

The procedure is straightforward and involves cross-validation to keep separate the data used for choosing which tests to perform and the data used for constructing the test statistics and generating p-values. Define the observed data as $O_i = (W_i, A_i, Y_i) \sim P_0, i=1,...,n$, which consist of $n$ independent and identically distributed random samples from the population distribution $P_0$, with $P_0$ known to be an element of a statistical model $\mathcal{M}$. For each observation, $Y_i$ is the outcome of interest of dimension $p$ and can be a real number or class variable. $A_i$ is a 1-dimensional vector of treatment variable and $W_i$ is a multi-dimensional vector of baseline covariates.

Consider  \textit{V-fold cross-validation}, where the learning set $L$ is randomly divided into $V$ mutually exclusive and exhaustive sets, $L_v, v=1, ... , V$, of as nearly equal size as possible. We will define the parameter-generating sample for $v, L - L_v$, the sample used to select which of the original set of hypotheses should be considered for future testing.  Then, the estimates are computed on the estimation-sample $L_v$, and averaged (details below) across the $v$. The proportion $p$ of observations in the estimation-sample $L_v$ is approximately $1/V$.

For a given random split $v$, let $P_{n,v^c}$ be the empirical distribution of the parameter-generating sample $L - L_v$, and $P_{n,v}$ be the empirical distribution of the estimation-sample $L_v$. ${\Psi _{{v},P_{n,v^c}}}:\mathcal{M} \to \mathbb{R} $ is the target
parameter mapping indexed by the parameter-generating sample $ P_{n,v^c}$,
and ${\widehat \Psi _{{v},P_{n,v^c}}}: {\mathcal{M}_{NP}} \to \mathbb{R}$
the corresponding estimator of this target parameter. For instance, one might
use an algorithm within the training sample to subset variables based on, say,
ranking by some statistic (e.g., differential expression).  Thus, ${\Psi_{{v},P_{n,v^c}}}$
could be the mean difference between two groups of this random subset of
variables. Here, $\mathcal{M}$ is a nonparametric model and an estimator is
defined as an algorithmic mapping from a nonparametric model, including all of
the constituent empirical distributions, to the parameter space. For
simplicity, assume that the parameter is real-valued. Thus, the target parameter
mapping and estimator can depend not only on the parameter-generating sample
${P_{n,v^c}}$, but also on the particular split $v$.

Assume the existence of a mapping from the parameter-generating sample
${P_{n,v^c}}$ into a target parameter mapping and a corresponding estimator
of that target parameter. The choice of target parameter mapping and
corresponding estimator can be informed by the data ${P_{n,v^c}}$ but not
by the estimation sample ${P_{n,v}}$ -- that is, one only need know the
realization of the mapping from the parameter-generating sample to the space of
target parameter mappings and estimators
$\left( {{\Psi_{{v},P_{n,v^c}}},{{\widehat
\Psi }_{{v},P_{n,v^c}}}} \right)$,
but not the explicit definition of said mapping.

Define the sample-split data-adaptive statistical target parameter as
${\Psi _n}:\mathcal{M} \to \mathbb{R}$ with

$$
{\Psi _n}\left( P \right) =
{E_{{v}}}{\Psi _{{v},P_{n,v^c}}}\left( P \right)
$$
and the statistical estimand of interest is thus
$$
{\Psi _{n,0}} = {\Psi _n}\left( P_0 \right) =
{E_{{v}}}{\Psi _{{v},P_{n,v^c}}}\left( {{P_0}} \right)=\underset{v}{Ave}\Psi _{{v},P_{n,v^c}}\left( P_0 \right).
$$

Note that this target parameter mapping depends on the data, which is the
reason for calling it a data-adaptive target parameter. A corresponding
estimator of the data-adaptive estimand ${\Psi _{n,0}}$ is given by:

$$
{\Psi _n} =  \Psi \left( {{P_n}} \right) =
{E_{{v}}}{ \Psi _{{v},P_{n,v^c}}}\left( {P_{n,v}} \right).
$$

\subsection{Data-adaptive test statistics}\label{stat}

In this section, we consider a rank-based approach to generate a test statistic on a reduced set of responses ${S^0} \subset Y$, where
$S^0$ is a subset of the full observed data determined by the application of a
selection procedure on the parameter-generating sample. The data-adaptive statistic is identified by three components: the cardinality of the reduced set denoted by ${p^*} \triangleq \left| {{S^0}} \right|$, the parameter-generating algorithm, and the number of folds $V$ in cross-validation.

Specifically, for each parameter-generating sample $L-L_v$, we simply rank  by ${\widehat B_{j}} = {E_{P_{n,v^c}}}\left( {\left. {{Y_j}}
\right|A = 1} \right) - {E_{P_{n,v^c}}}\left( {\left. {{Y_j}} \right|A = 0}
\right)$, which is the empirical average treatment effect of $A$ on $Y_j$. Within each fold $v$, the parameter-generating algorithm returns the rank
of each response covariate by its effect size $\widehat B_j$ using the the parameter-generating sample $L-L_v$, and the set
${S_{P_{n,v^c}}}$ is defined by taking the top $p^*$ $j$'s (
we have arbitrarily chosen $p^* = 30$).

The parameter of interest is the average treatment effect ${\Psi _{n,j}} = {E_{{P_n}}}\left( {\left. {{Y_j}} \right|A = 1} \right) - {E_{{P_n}}}\left( {\left. {{Y_j}} \right|A = 0} \right)$, estimated by ${\Psi _{{v},P_{n,v^c},j}}\left( {P_{n,v}} \right) = {E_{{P_{n,v}}}}\left( {\left. {{Y_{{S_{{P_{n,{v^c}}}}},j}}} \right|A = 1} \right) - {E_{{P_{n,v}}}}\left( {\left. {{Y_{{S_{{P_{n,{v^c}}}}},j}}} \right|A = 0} \right)$ using the estimation-sample $L_v$. The efficient influence curve of $ \Psi _{{v},P_{n,v^c},j}$ is derived in \cite{van2012tmle} and can be represented as $${D^*}(P)(O) = H(g)(A,W)\left( {Y - \overline Q (A,W)} \right) + \overline Q (1,W) - \overline Q (0,W) - \Psi (Q),$$ where $\overline Q (a,W) = {E_P}(Y|W,A = a)$, $H(g)(A,W) = (2A - 1)/g(A|W)$, and $g(A|W) = {E_P}(A = 1|W)$. We calculate the efficient influence curves ${D^*}(P_{n,v^c})(O)$ also on the estimation-sample, which will be useful when we later calculate test statistics.

We repeat the procedure for all $V$ folds, and take the average $ {E_{{v}}}{ \Psi _{{v},P_{n,v^c,j}}}\left( {P_{n,v}} \right)$ for each $Y_j$ that are selected in any single fold $S_{P_{n,v^c}}, v=1,…,V$. The calculated efficient influence curves are combined across all $V$ folds and used to derive asymptotic distribution of target parameters and perform statistical testing. The p-values for the $Y_j$ in the final reduced set can be constructed based on asymptotic linearity of the TMLE \citep{van2012tmle}. Under regularity conditions on the estimates of $Q (A,W)$ and $ g(A|W)$,
$$\sqrt n \left( {{\psi _{n,{P_n},j}}({P_n}) - {\psi _{n,{P_n},j}}({P_0})} \right)\sim N\left( {0,{\sigma _{EIC}}^2} \right),$$
where $${\sigma _{EIC,j}}^2 = \frac{1}{n}\sum\limits_{i = 1}^n {{D^*}{{_j}^2}({O_i})} $$ is the empirical variance of efficient influence curve. As a result, the test statistics based on the response matrix $Y_{S^0}$ are computed using asymptotic normal distribution of each single target parameter, thus we have a p-value for each selected $Y_j$ in $S^0$. The false discovery rate of the corresponding test p-values
(${p_1}, \dots ,{p_{{p^*}}}$) can be controlled, for example, using the
well-established Benjamini-Hochberg procedure for controlling the False
Discovery Rate (FDR) \citep{benjamini1995controlling}.

The data-adaptive parameter-generating procedure not only reduces the multiplicity of the hypothesis tests compared with directly applying multiple testing methodologies, but it also generates summary statistics that validate the robustness and credibility of the result. A plot of sorted Q-values can become useful when a lot of covariates are significant. If the smallest Q-values are similar among each other, and as a group much smaller than the rest (still significant) covariates, we can identify the cluster of smallest Q-values to be of most scientific interest. In addition to looking at the Q-values, which is a measure of statistical significance, practitioners can evaluate the scientific significance by looking at the magnitude of average treatment effect. Percentage of each $Y_j$ to be selected in $S^0_v$ across all $V$ folds can be viewed as a measure of association robustness. So is the average rank of each covariate across $V$ folds.

\newpage

\section{Simulation}\label{sim}

We consider a situation that has an analogue in high-dimensional data generated
by  microRNA data discussed above. The method evaluated in this simulation
study uses the parameter-generating sample to select a small subset of the
original genes, and subsequently it uses the estimation sample to validate the
effect of these genes on a phenotype of interest. In this manner, it avoids the
need to apply multiple testing procedures that control a Type-I error rate
among a comparatively large number of tests.

Let $O = \left( {A,Y = \left( {{Y_1},{Y_2}, \dots ,{Y_p}} \right)} \right)$
where $A$ is a binary vector, and $Y$ a multivariate outcome. The true
probability distribution, $P_0$, is generated based on a design where there is
equal probability of $A = 0$ and $A = 1$; moreover, for each gene $j$, the
distribution of $Y_j$, given $A$, is defined by the following regression
equation:

$${Y_j} = {B_{0,j}} + {B_{1,j}}A + {e_j},\: j = 1, \dots, p$$

The coefficient ${B_{0,j}}$ is generated by a standard normal distribution, and
the coefficient ${B_{1,j}}$ takes a fixed sparse design, with
${B_{1,1}} = {B_{1,2}} = \dots = {B_{1,10}} = 1$ and
${B_{1,11}} = {B_{1,12}} = \dots = {B_{1,p}} = 0$. As a result, we generated
$10$ true effects and $(p-10)$ null effects. Note, that these coefficients are
fixed in the simulation. The errors $e_j$ were independent draws from a random
$N\left( {0,\sigma _e^2} \right)$ distribution, and we repeated the simulation
not only for different magnitudes of the residual error (different realizations
of ${\sigma _e^2}$), but also for increasing sample sizes. We define our
data-adaptive statistical target parameter as outlined in
\ref{notation}, where we set the dimension of the reduced response
matrix $p^*$ to be $15$. Data on a total of $p = 10^6$ potential biomarkers were generated.

\begin{figure}[h!]
    \centering
    \includegraphics[width=0.75\textwidth]{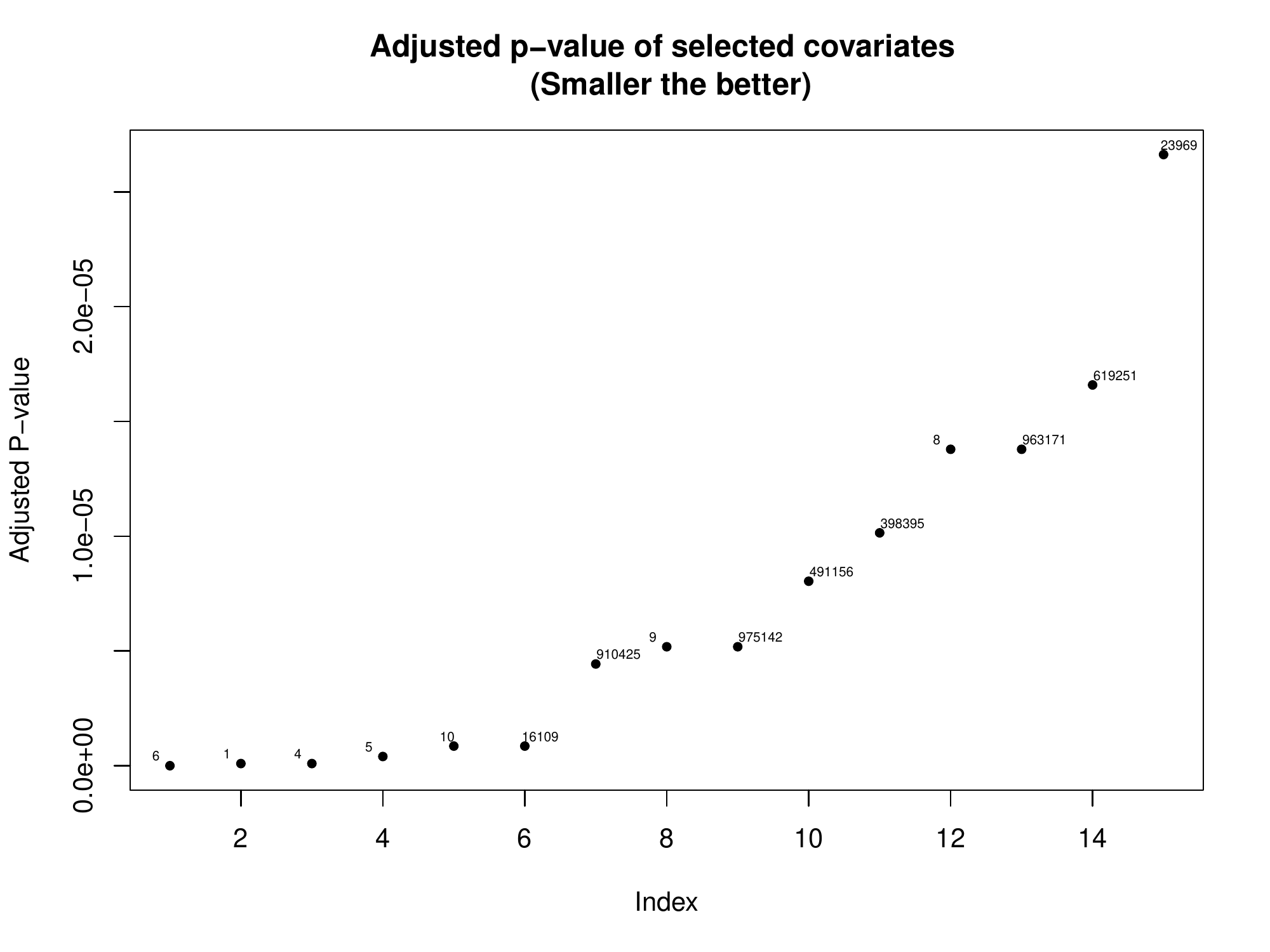}
    \caption{Plot of adjusted p-values using data-adaptive test statistics}
    \label{fig:adaptBH}
\end{figure}

Directly adjusting p-values using the method of
\cite{benjamini1995controlling}, for controlling the FDR, on all $10^6$
responses yields the plot in \ref{fig:adaptBH}. Note that $8$ out of the top
$10$  true effects failed to achieve significance despite a signal-to-noise
ratio of $10$, due to adjustment on too large a dimension.

\begin{figure}[!h]
    \centering
    \includegraphics[width=0.75\textwidth]{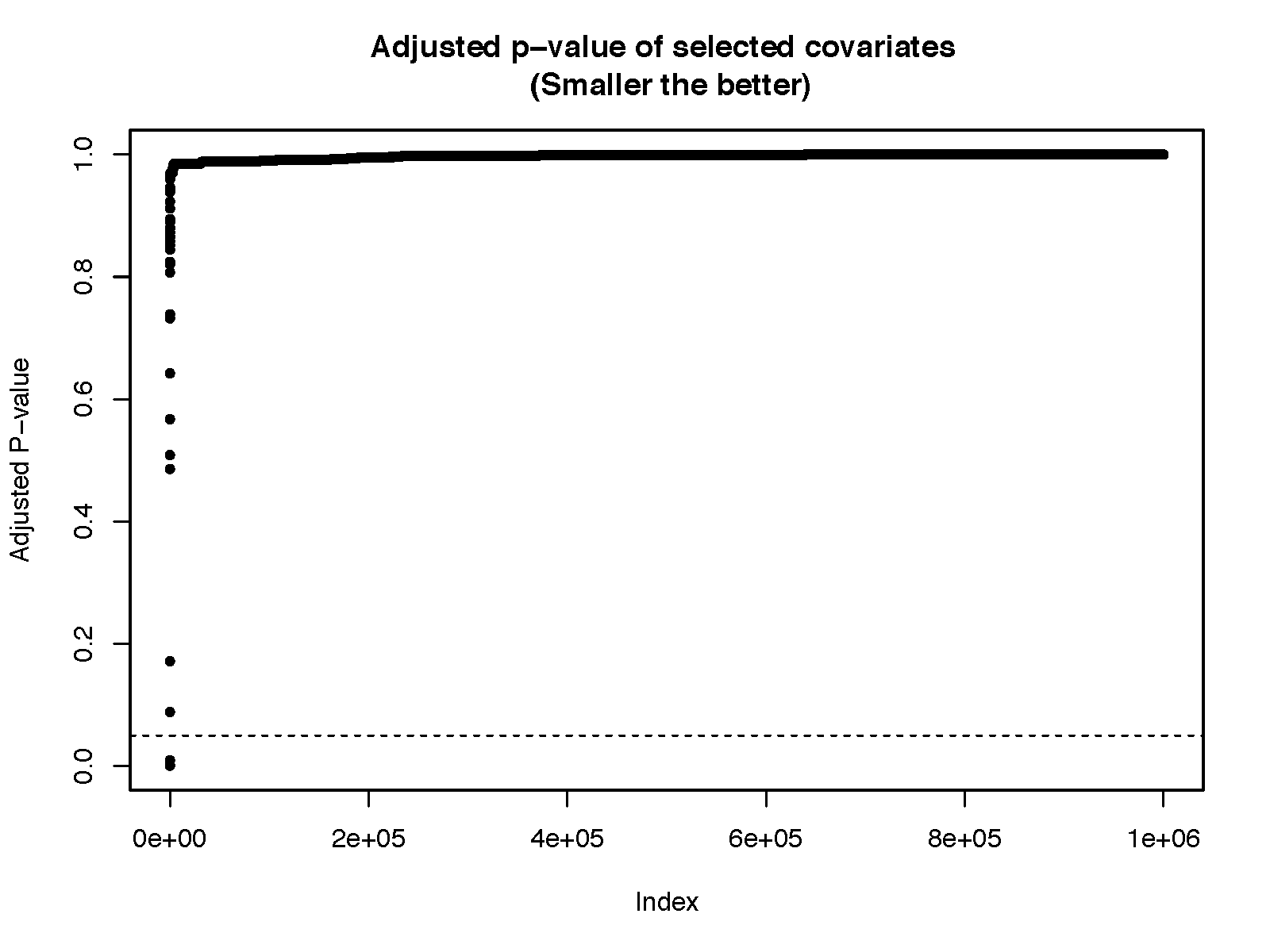}
    \caption{Plot of adjusted p-values using all response covariates}
\end{figure}

Running the data-adaptive algorithm on the parameter-generating samples
provides a reasonable recovery of true effect responses. The results are given
below:

\begin{table}[h!]
\centering
\caption{Summary of the results of data-adaptive statistical target parameter
estimation}
\begin{tabular}{lllllll}
\hline
\hline
& covar. ID & ATE est. & p-value & adjusted p-value & mean CV-rank & \% times
in top 15 \\
\hline
1 & 6 & 1.3474 & 2.05E-10 & 3.07E-09 & 1.1 & 100 \\
2 & 1 & 1.2498 & 1.35E-08 & 9.72E-08 & 2.4 & 100 \\
3 & 4 & 1.2406 & 1.94E-08 & 9.72E-08 & 2.8 & 100 \\
4 & 5 & 1.0822 & 1.08E-07 & 4.04E-07 & 4.8 & 100 \\
5 & 10 & 1.0250 & 3.43E-07 & 8.58E-07 & 7.3 & 90 \\
6 & 16109 & 0.9605 & 3.32E-07 & 8.58E-07 & 11.4 & 70 \\
7 & 8 & 0.9793 & 1.12E-05 & 1.38E-05 & 11.5 & 70 \\
8 & 9 & 0.9073 & 3.11E-06 & 5.18E-06 & 26.9 & 40 \\
9 & 23969 & 0.9172 & 2.66E-05 & 2.66E-05 & 27 & 50 \\
10 & 910425 & 0.9107 & 2.07E-06 & 4.43E-06 & 27.1 & 30 \\
11 & 398395 & 0.8975 & 7.44E-06 & 1.01E-05 & 27.7 & 20 \\
12 & 975142 & 0.9127 & 2.94E-06 & 5.18E-06 & 28.1 & 40 \\
13 & 963171 & 0.9124 & 1.19E-05 & 1.38E-05 & 28.7 & 40 \\
14 & 491156 & 0.9425 & 5.36E-06 & 8.04E-06 & 31.7 & 50 \\
15 & 619251 & 0.8970 & 1.55E-05 & 1.66E-05 & 35.2 & 50 \\
\hline
\end{tabular}
\end{table}

From the table, it is clear that the approach of data-adaptive statistical
target parameter estimation consistently picks out the true effects in the top
candidates. After application of the \cite{benjamini1995controlling} procedure
on the reduced set of responses, $74$ out of $10$ true signals are still
significant.

\newpage

\section{Differentially expressed microRNA and exposure to benzene}\label{mirna}

Benzene, an established cause of acute myeloid leukemia (AML), may also cause
one or more lymphoid malignancies in humans. Previous studies have identified
single nucleotide polymorphisms (SNP) and patterns of DNA methylation
associated with exposure to benzene through transcriptomic analyses of blood
cells from a small number of occupationally exposed workers
\citep{zhang2010systems}; however, in these studies, the effect of
benzene-induced changes to microRNA expression (Affymetrix 3.0 GeneChips which
contain probes for $5639$ human miRNAs) in blood cells was not the
subject of intense scrutiny. We discuss a study of 85 individuals in Tianjin,
China, in which 56 workers exposed to varying levels of benzene and 29
unexposed control counterparts were monitored repeatedly for up to 12 months. For each individual, blood samples were collected and miRNA
expression was measured and log-transformed, leading to the formulation of a statistical problem in
which a large number of comparisons arises from performing miRNA screens on
both exposed subjects and unexposed controls.

To illustrate the flexibility of the proposed method we will show how it can be used as a powerful test for differentially expressed microRNA.
The data consist of 5639 real-valued outcomes and one binary treatment for 85 individuals. In univariate testing, the microRNA hsa-miR-320a\_st has the smallest p-value ($p < 0.000455$).
However the association is no longer significant after controlling for multiple testing using the Benjamini-Hochberg \citep{benjamini1995controlling} method to control the False Discovery Rate ($p_{adjusted} = 0.5281$).
Since our objective of interest was to detect the top differentially expressed microRNAs.
(and not to differentiate all microRNAs), we generated data-adaptive test statistics to reduce the number of hypotheses based on the procedure outlined in \ref{notation}

\begin{table}[ht]
\centering
\caption{Top microRNA's after direct applying Benjamini-Hochberg (Down-regulated)}
\label{ttest_nega}
\begin{tabular}{rlrrrr}
  \hline
 & microRNA & ATE & Fold Change & p-value & adjusted p-value \\
  \hline
1 & hsa-miR-744\_st & -0.35 & 0.79 & 0.00067 & 0.52814 \\
  2 & hsa-miR-320b\_st & -0.32 & 0.80 & 0.00046 & 0.52814 \\
  3 & hsa-miR-320a\_st & -0.31 & 0.80 & 0.00045 & 0.52814 \\
  4 & hsa-miR-320c\_st & -0.30 & 0.81 & 0.00082 & 0.52814 \\
  5 & hp\_hsa-mir-449b\_x\_st & -0.23 & 0.85 & 0.00084 & 0.52814 \\
  6 & ENSG00000238375\_st & -0.20 & 0.87 & 0.00071 & 0.52814 \\
  7 & hp\_hsa-mir-4645\_st & -0.18 & 0.88 & 0.00077 & 0.52814 \\
   \hline
\end{tabular}
\end{table}

\begin{table}[ht]
\centering
\caption{Top microRNA's after direct applying Benjamini-Hochberg (Up-regulated)}
\label{ttest_posi}
\begin{tabular}{rlrrrr}
  \hline
 & microRNA & ATE & Fold Change & p-value & adjusted p-value \\
  \hline
1 & hsa-miR-338-5p\_st & 0.83 & 1.78 & 0.00079 & 0.52814 \\
 2 & hsa-miR-103a-2-star\_st & 0.66 & 1.58 & 0.00061 & 0.52814 \\
 3 & hsa-miR-4725-3p\_st & 0.34 & 1.26 & 0.00094 & 0.53127 \\
   \hline
\end{tabular}
\end{table}

\begin{figure}[h!]
    \centering
    (a)\includegraphics[width=0.45\textwidth]{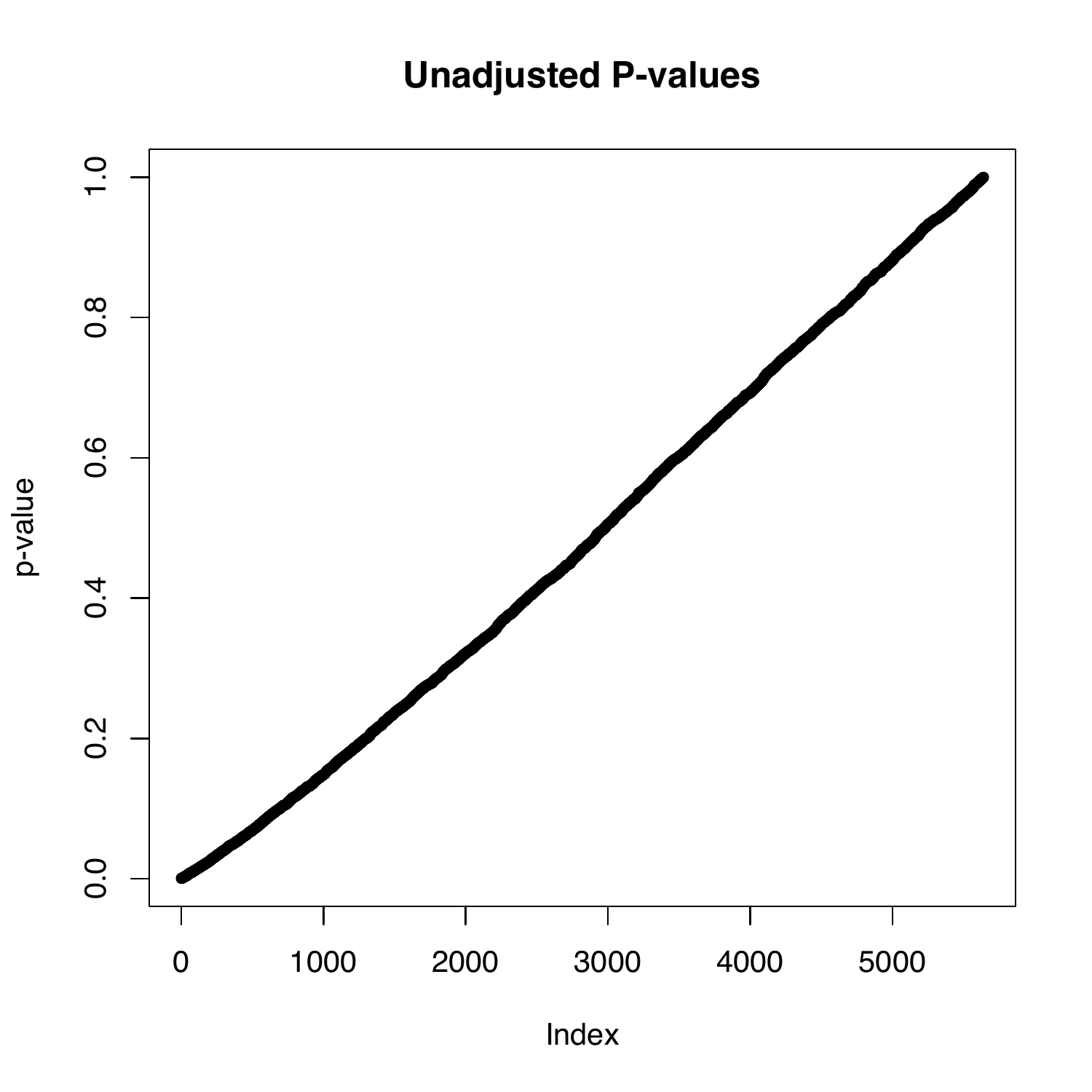}
    (b)\includegraphics[width=0.45\textwidth]{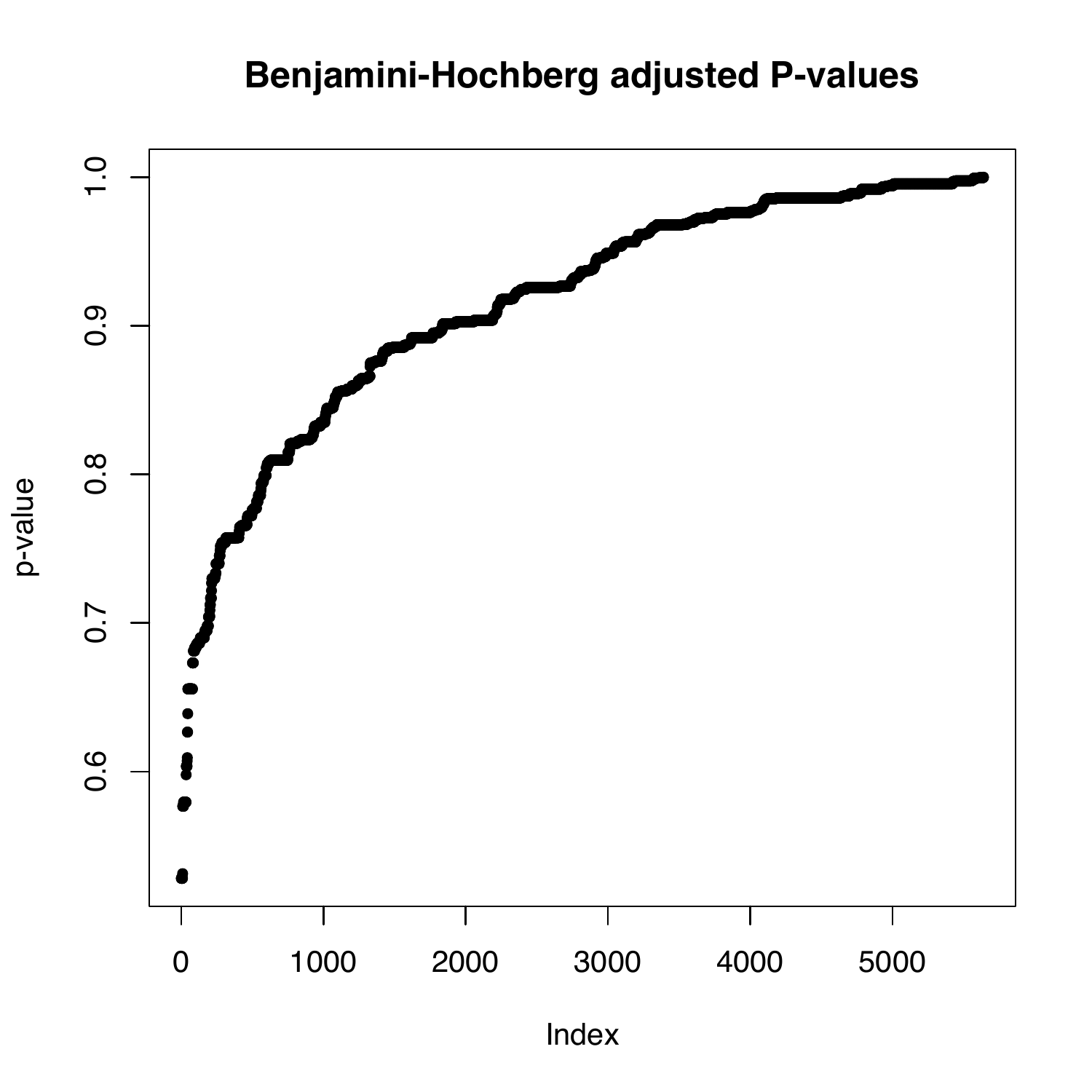}
    \caption{Plot of (a) unadjusted and (b) FDR adjusted p-values using all covariates}
    \label{fig:adjPvalsBH}
\end{figure}

Given the multiplicity of comparisons, we propose the use of data-adaptive test statistics to reduce the number of comparisons first, thereby increasing power, while still maintaining
accurate statistical inference. We specified the reduced set of response matrix $\widehat Y$ with dimension $n \times 30$, which correspond to studying top 30 microRNAs that will express differently under benzene exposure. We carried out 10-fold cross-validation to calculate the data-adaptive test statistics and tested each of the top 30 microRNAs. We finally performed FDR correction (\citep{benjamini1995controlling}) on the 30 raw p-values.

\subsection{Results}

\begin{figure}[!htbp]
    \centering
    \includegraphics[width=0.75\textwidth]{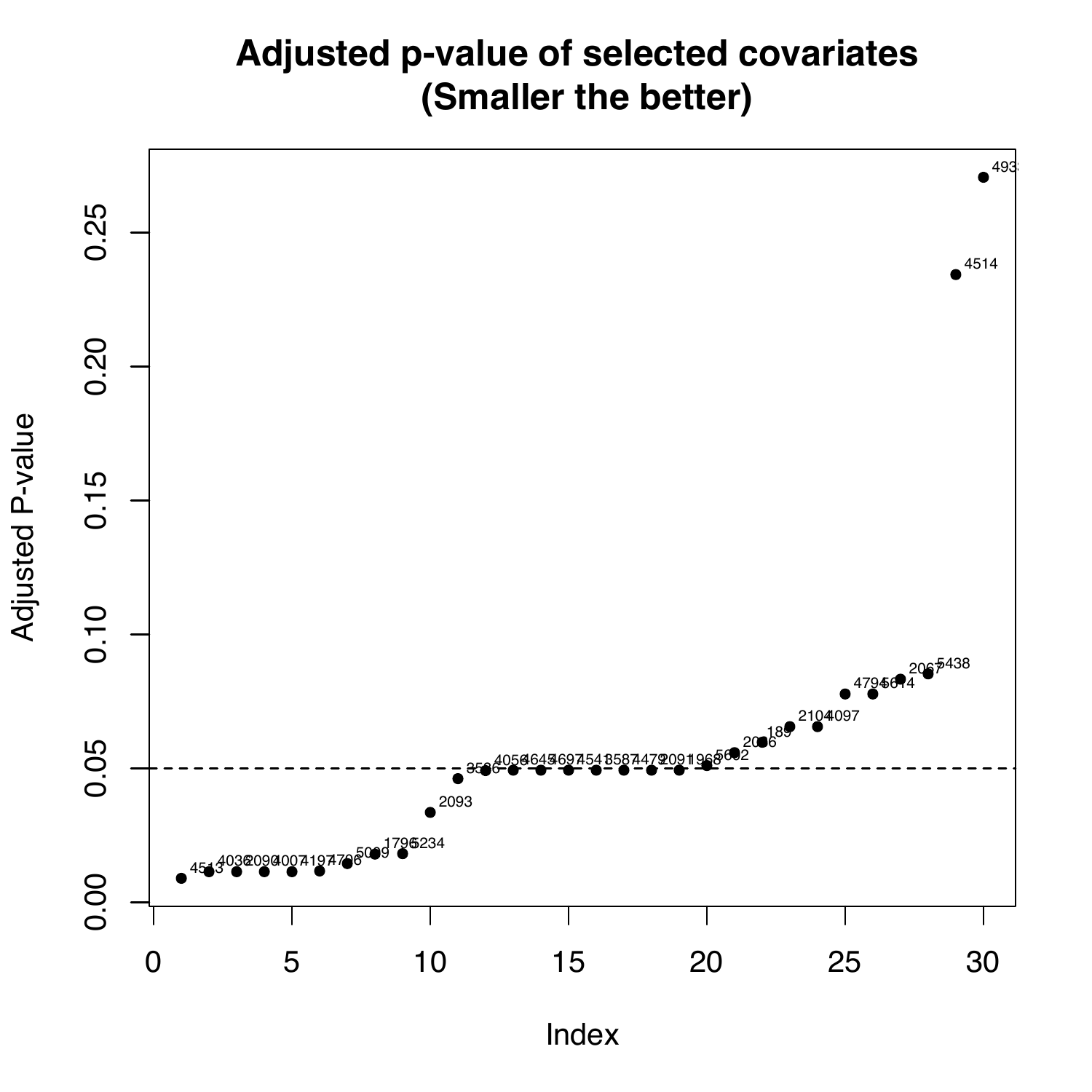}
    \caption{Adjusted p-values using data-adaptive test statistics (Down-regulated)}
    \label{fig:adjPvalsDataAdapt_nega}
\end{figure}

The results for the top 30 microRNAs and their FDR-adjusted p-values are shown in Table \ref{adapt_nega}. 19 out of 30 top microRNAs had a significant differential expression (q-value $< 0.05$) while we found none in FDR-corrected t-tests in Table \ref{ttest_nega}. Observing the plot of FDR-adjusted p-values (Figure \ref{fig:adjPvalsDataAdapt_nega}) also gives us insights as we can easily identify groups of significant p-values. In practice, we can choose a cutoff based on the trend of sorted p-values. The average rank of the top covariates (in Table \ref{adapt_nega}) can also be referenced as a measure of stability of the effect across different subjects.

\begin{table}[!htbp]
\centering
\caption{Summary of the data-adaptive test statistics on miRNA data (Down-regulated)}
\label{adapt_nega}
\begin{tabular}{rlrrrrr}
  \hline
 & microRNA & ATE & raw p-values & adjusted p-values & avg rank & \% appear in top 30 \\
  \hline
1 & hsa-miR-134\_st & -0.87 & 0.0197 & 0.0492 & 2.4 & 100 \\
  2 & hsa-miR-3613-3p\_st & -0.86 & 0.0003 & 0.0089 & 2.7 & 100 \\
  3 & hsa-miR-4668-5p\_st & -0.77 & 0.0034 & 0.0144 & 4.5 & 100 \\
  4 & hsa-miR-382\_st & -0.80 & 0.0294 & 0.0493 & 6.2 & 100 \\
  5 & U49A\_s\_st & -0.70 & 0.0019 & 0.0114 & 7.5 & 100 \\
  6 & hsa-miR-409-3p\_st & -0.75 & 0.0312 & 0.0493 & 8.1 & 100 \\
  7 & hsa-miR-3651\_st & -0.67 & 0.0271 & 0.0493 & 8.7 & 100 \\
  8 & hsa-miR-432\_st & -0.72 & 0.0657 & 0.0777 & 10.0 & 100 \\
  9 & hp\_hsa-mir-548ai\_st & -0.63 & 0.0169 & 0.0461 & 11.6 & 100 \\
  10 & hsa-miR-1301\_st & -0.61 & 0.0008 & 0.0114 & 11.7 & 100 \\
  11 & hsa-miR-1275\_st & -0.57 & 0.0015 & 0.0114 & 14.9 & 100 \\
  12 & hsa-miR-200c\_st & -0.57 & 0.0016 & 0.0114 & 16.2 & 90 \\
  13 & ENSG00000199411\_s\_st & -0.56 & 0.0438 & 0.0597 & 16.7 & 80 \\
  14 & hp\_hsa-mir-548ai\_x\_st & -0.52 & 0.0229 & 0.0493 & 22.8 & 80 \\
  15 & hsa-miR-423-5p\_st & -0.50 & 0.0023 & 0.0117 & 23.6 & 80 \\
  16 & U56\_st & -0.51 & 0.0503 & 0.0656 & 24.5 & 80 \\
  17 & ENSG00000252921\_x\_st & -0.49 & 0.0048 & 0.0180 & 26.9 & 70 \\
  18 & U49B\_s\_st & -0.49 & 0.0112 & 0.0335 & 27.3 & 80 \\
  19 & U38A\_st & -0.49 & 0.0749 & 0.0833 & 28.4 & 70 \\
  20 & hsa-miR-3613-5p\_st & -0.51 & 0.2265 & 0.2343 & 30.5 & 70 \\
  21 & hsa-miR-99b\_st & -0.47 & 0.0674 & 0.0777 & 33.8 & 50 \\
  22 & hsa-miR-486-5p\_st & -0.46 & 0.0054 & 0.0181 & 34.5 & 70 \\
  23 & hsa-miR-339-3p\_st & -0.45 & 0.0245 & 0.0493 & 37.3 & 40 \\
  24 & U49A\_st & -0.43 & 0.0309 & 0.0493 & 39.7 & 30 \\
  25 & HBII-85-2\_x\_st & -0.43 & 0.0247 & 0.0493 & 39.9 & 20 \\
  26 & U21\_st & -0.44 & 0.0391 & 0.0559 & 39.9 & 40 \\
  27 & hsa-miR-4529-3p\_st & -0.49 & 0.2706 & 0.2706 & 42.0 & 50 \\
  28 & hsa-miR-940\_st & -0.44 & 0.0340 & 0.0510 & 42.2 & 50 \\
  29 & hsa-miR-584\_st & -0.45 & 0.0796 & 0.0853 & 42.3 & 50 \\
  30 & hsa-miR-150-star\_st & -0.43 & 0.0524 & 0.0656 & 43.6 & 20 \\
   \hline
\end{tabular}
\end{table}

\begin{table}[!htbp]
\centering
\caption{List of microRNA's that are still significant after FDR correction (Down-regulated)}
\begin{tabular}{rl}
  \hline
 & microRNA \\
  \hline
1 & hsa-miR-134\_st \\
  2 & hsa-miR-3613-3p\_st \\
  3 & hsa-miR-4668-5p\_st \\
  4 & hsa-miR-382\_st \\
  5 & U49A\_s\_st \\
  6 & hsa-miR-409-3p\_st \\
  7 & hsa-miR-3651\_st \\
  8 & hp\_hsa-mir-548ai\_st \\
  9 & hsa-miR-1301\_st \\
  10 & hsa-miR-1275\_st \\
  11 & hsa-miR-200c\_st \\
  12 & hp\_hsa-mir-548ai\_x\_st \\
  13 & hsa-miR-423-5p\_st \\
  14 & ENSG00000252921\_x\_st \\
  15 & U49B\_s\_st \\
  16 & hsa-miR-486-5p\_st \\
  17 & hsa-miR-339-3p\_st \\
  18 & U49A\_st \\
  19 & HBII-85-2\_x\_st \\
   \hline
\end{tabular}
\end{table}

\newpage

The same analysis can be performed on up-regulated microRNA's. 20 out of 30 top microRNAs had a significant differential expression (q-value $< 0.05$) as did the hsa-miR-338-5p\_st and hsa-miR-103a-2-star\_st that we found in FDR-corrected t-tests in Table \ref{ttest_posi}.


\begin{table}[!htbp]
\centering
\caption{Summary of the data-adaptive test statistics on miRNA data (Up-regulated)}
\label{adapt_posi}
\begin{tabular}{lllllll}
\hline
\hline
   & microRNA                & ATE & raw p-value & adjusted p-value & avg rank & \% appear in top 30 \\
\hline
1  & hsa-miR-505\_st         & 1.0175       & 0.001144    & 0.006465         & 1.2             & 100                 \\
2  & hsa-miR-4772-3p\_st     & 0.8763       & 0.001508    & 0.006465         & 3.4             & 100                 \\
3  & hsa-miR-10a\_st         & 0.8983       & 0.000979    & 0.006465         & 3.8             & 100                 \\
4  & hsa-miR-338-5p\_st      & 0.8313       & 0.000155    & 0.002986         & 5.3             & 100                 \\
5  & hsa-miR-301a\_st        & 0.8373       & 0.011351    & 0.026194         & 5.4             & 100                 \\
6  & hsa-miR-212\_st         & 0.8158       & 0.001421    & 0.006465         & 6.1             & 100                 \\
7  & hsa-miR-374b\_st        & 0.7728       & 0.035704    & 0.051006         & 7.6             & 100                 \\
8  & hsa-miR-454\_st         & 0.776        & 0.012762    & 0.02691          & 7.9             & 100                 \\
9  & hsa-miR-7-1-star\_st    & 0.7271       & 0.013911    & 0.02691          & 10.1            & 100                 \\
10 & hsa-miR-4674\_st        & 0.727        & 0.047842    & 0.064645         & 12.4            & 90                  \\
11 & hsa-miR-30b\_st         & 0.6837       & 0.094063    & 0.104514         & 14.3            & 100                 \\
12 & hsa-miR-29c-star\_st    & 0.6499       & 0.002409    & 0.009034         & 15.5            & 100                 \\
13 & hsa-miR-103a-2-star\_st & 0.6597       & 0.000199    & 0.002986         & 15.8            & 100                 \\
14 & hsa-let-7d-star\_st     & 0.6519       & 0.014352    & 0.02691          & 16.4            & 100                 \\
15 & hsa-miR-142-5p\_st      & 0.6668       & 0.049561    & 0.064645         & 17.7            & 90                  \\
16 & hsa-miR-361-3p\_st      & 0.6136       & 0.02849     & 0.043226         & 20.5            & 90                  \\
17 & hsa-miR-1231\_st        & 0.6032       & 0.005194    & 0.01501          & 21.1            & 90                  \\
18 & hsa-miR-99a\_st         & 0.6099       & 0.119402    & 0.123519         & 22.4            & 80                  \\
19 & hsa-miR-589-star\_st    & 0.5867       & 0.018542    & 0.032722         & 23              & 80                  \\
20 & hsa-miR-3188\_st        & 0.581        & 0.001158    & 0.006465         & 24.3            & 80                  \\
21 & hsa-miR-3621\_st        & 0.5967       & 0.091534    & 0.104514         & 25              & 70                  \\
22 & hsa-let-7g\_st          & 0.5883       & 0.212763    & 0.212763         & 29.3            & 70                  \\
23 & hsa-miR-148a\_st        & 0.5586       & 0.113839    & 0.12197          & 29.3            & 60                  \\
24 & hsa-miR-378g\_st        & 0.537        & 0.021924    & 0.036541         & 32.1            & 40                  \\
25 & hsa-miR-30e-star\_st    & 0.5392       & 0.082371    & 0.098846         & 32.6            & 40                  \\
26 & hsa-miR-641\_st         & 0.531        & 0.005503    & 0.01501          & 33.3            & 30                  \\
27 & hsa-miR-221-star\_st    & 0.546        & 0.028817    & 0.043226         & 33.6            & 60                  \\
28 & hsa-miR-181a-star\_st   & 0.5398       & 0.057202    & 0.071502         & 34.1            & 60                  \\
29 & hsa-miR-186\_st         & 0.5304       & 0.009065    & 0.02266          & 34.5            & 30                  \\
30 & hsa-miR-3187-3p\_st     & 0.524        & 0.005102    & 0.01501          & 35.7            & 30 \\
\hline

\end{tabular}
\end{table}

\begin{table}[!htbp]
\centering
\caption{List of microRNA's that are still significant after FDR correction (Up-regulated)}
\begin{tabular}{rl}
  \hline
 & microRNA \\
  \hline
1 & hsa-miR-505\_st \\
  2 & hsa-miR-4772-3p\_st \\
  3 & hsa-miR-10a\_st \\
  4 & hsa-miR-338-5p\_st \\
  5 & hsa-miR-301a\_st \\
  6 & hsa-miR-212\_st \\
  7 & hsa-miR-454\_st \\
  8 & hsa-miR-7-1-star\_st \\
  9 & hsa-miR-29c-star\_st \\
  10 & hsa-miR-103a-2-star\_st \\
  11 & hsa-let-7d-star\_st \\
  12 & hsa-miR-361-3p\_st \\
  13 & hsa-miR-1231\_st \\
  14 & hsa-miR-589-star\_st \\
  15 & hsa-miR-3188\_st \\
  16 & hsa-miR-378g\_st \\
  17 & hsa-miR-641\_st \\
  18 & hsa-miR-221-star\_st \\
  19 & hsa-miR-186\_st \\
  20 & hsa-miR-3187-3p\_st \\
   \hline
\end{tabular}
\end{table}

\begin{figure}[!htbp]
    \centering
    \includegraphics[width=0.75\textwidth]{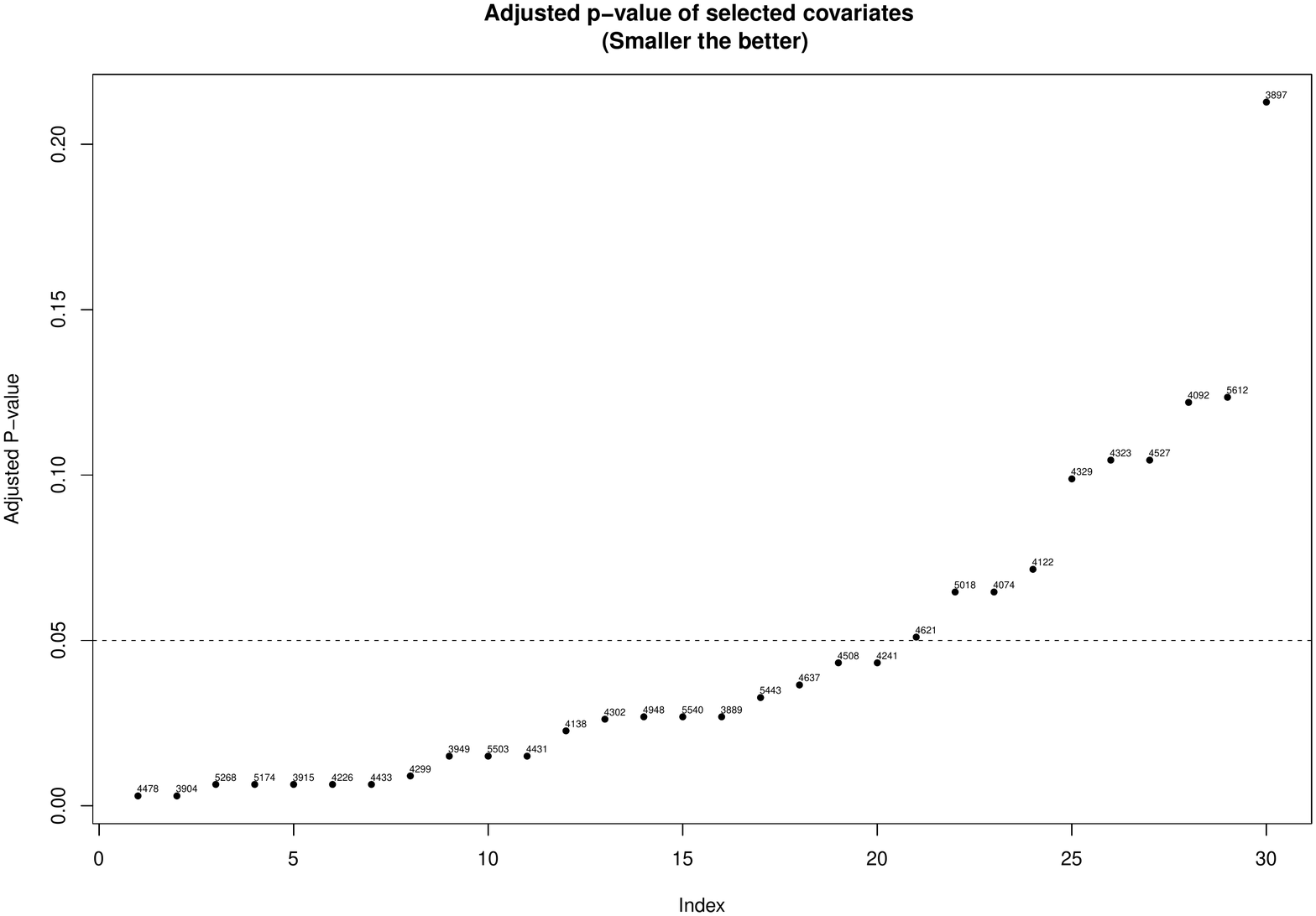}
    \caption{Adjusted p-values using data-adaptive test statistics (Up-regulated)}
    \label{fig:adjPvalsDataAdapt_posi}
\end{figure}


\newpage
Overall, this formally confirms the conclusions in the paper that by generating the data-adaptive test statistics we can increase the power of testing a large set of statistical hypotheses and at the same time control the level of false positive rate (or false discovery rate).

\newpage

\section{Discussion}\label{discussion}

The goal of this article is to introduce a generalized class of robust
procedures for performing statistical tests in high-dimensional settings,
relying on the approach of data-adaptive statistical target parameters. Here,
we have introduced, in generality, the theory and methodology underlying the use
of data-adaptive statistics for multiple testing, illustrating key advantages of
this approach via simulation studies and providing examples where relevant. By
providing the theoretical formalisms in a generalized way, we have exposed a
flexible framework in which the number of multiple testing corrections applied
in high-dimensional problems can be reduced, allowing for signals that would
otherwise be made undetectable by said corrections to be recovered.

In the example provided, we demonstrate the power of the approach based on
data-adaptive test statistics in the context of a study of miRNA. We show that
this new class of approaches for analyzing high-dimensional data sets allows
researchers to derive improved statistical power in problems plagued by
multiple testing, by allowing for relatively fewer null hypotheses of interest
to be generated data-adaptively -- that is, suggested by the observed data. In
order to improve accessibilty to the methodology presented herein,
\cite{cai2017data} have developed and made publicly available an open-source
software package for data-adaptive multiple testing, available for the R
statistical computing language (\cite{R}).



\newpage

\nocite{*}
\bibliographystyle{elsarticle-harv}
\bibliography{manuscript}

\end{document}